# Publication Bias (The "File-Drawer Problem") in Scientific Inference

Jeffrey D. Scargle

Space Science Division

National Aeronautics and Space Administration

Ames Research Center

jeffrey@sunshine.arc.nasa.gov

In honor of Professor Peter A. Sturrock

Presented at the

**Sturrock Symposium**

Saturday, March 20, 1999

Stanford University

# The "File-Drawer" Problem in Scientific Inference

Jeffrey D. Scargle

Space Science Division, NASA Ames Research Center

MS 245-3, Moffett Field, CA 94035-1000

`jeffrey@sunshine.arc.nasa.gov`

It is human nature for "the affirmative or active to effect more than the negative or privative. So that a few times hitting, or presence, countervails oft-times failing or absence."

... Francis Bacon, *The Advancement of Learning*

Received _______________; accepted _______________

## ABSTRACT

*Publication bias* arises whenever the probability that a study is published depends on the statistical significance of its results. This bias, often called the *file-drawer effect* since the unpublished results are imagined to be tucked away in researchers' file cabinets, is potentially a severe impediment to combining the statistical results of studies collected from the literature. With almost any reasonable quantitative model for publication bias, only a small number of studies lost in the file-drawer will produce a significant bias. This result contradicts the well known Fail Safe File Drawer (**FSFD**) method for setting limits on the potential harm of publication bias, widely used in social, medical and psychic research. This method incorrectly treats the file drawer as unbiased, and almost always misestimates the seriousness of publication bias. A large body of not only psychic research, but medical and social science studies, has mistakenly relied on this method to validate claimed discoveries. Statistical combination can be trusted only if it is known with certainty that all studies that have been carried out are included. Such certainty is virtually impossible to achieve in literature surveys.

# Contents

## 1. Introduction: Combined Studies

The goal of many studies in science, medicine, and engineering is the measurement of a quantity in order to detect a suspected effect or gain information about a known one. Observational errors and other noise sources make this a statistical endeavor, in which one obtains repeated measurements in order to average out these fluctuations.

If individual studies are not conclusive, improvement is possible by combining the results of different measurements of the same effect. The idea is to perform statistical analysis on relevant data collected from the literature[1] in order to improve the signal-to-noise (on the assumption that the noise averages to zero). A good overview of this topic can be found in (Hedges and Olkin 1985). For clarity and consistency with most of the literature, throughout this paper individually published results will be called *studies*, and the term *analysis* will refer to the endeavor to combine two or more studies.

Two problems arise in such analyses. First, experimenters often publish only statistical summaries of their studies, not the actual data. The analyst is then faced with combining the summaries, a non-trivial technical problem (see *e.g.* Rosenthal, 1978, Rosenthal, 1995). Modern communication technology should circumvent these problems by making even large data arrays accessible to other researchers. *Reproducible research* (Claerbout, 1999, Buckheit and Donoho, 1995) is a discipline for doing this and more, but even this methodology does not solve

---

[1]In some fields this is called *meta-analysis*, from the Greek *meta*, meaning *behind, after, higher or beyond*, and often denoting *change*. It usage here presumably refers to new issues arising in the statistical analysis of combined data – such as the file drawer effect itself. It is used mostly in scientific terminology, to imply a kind of superior or oversight status – as in *metaphysics*. I prefer the more straightforward term *combined analysis*.

the other problem, publication bias.

## 2.   Publication Bias

The second problem facing combined analysis is that studies collected from the literature are often not a representative sample. Erroneous statistical conclusions may result from a prejudiced collection process or if the literature is itself a biased sample of all relevant studies. The latter, *publication bias*, is the subject of this paper. The bibliography contains a sample of the rather large literature on the file drawer effect and publication bias. The essence of this effect, as described by nearly all authors, can be expressed in statistical language as follows:

<div style="border:1px solid black">

Definition: A **publication bias** exists if the probability that a study reaches the literature, and is thus available for combined analysis, depends on the results of the study.

</div>

What matters is whether the experimental results are actually used in the combined analysis, not just the question of publication. That is, the relevant process is this entire sequence, following the initiation of the study:

1. the study is carried out to a pre-defined stop point

2. all data are permanently recorded

3. data are reduced and analyzed

4. paper is written

5. paper is submitted to a journal

6. journal agrees to consider the paper

7. referee and author negotiate revisions

8. referee accepts paper

9. editor accepts paper for publication

10. author still wishes to publish paper

11. author's institution agrees to pay page charges

12. paper is published

13. paper is located during literature search

14. data from paper included in combined analysis

I refer to this concatenated process loosely as *publication*, but it is obviously more complex than what is usually meant by the term. Some of the steps may seem trivial, but all are relevant. Each step involves human decisions and so may be influenced by the result of the study. Publication probability may depend on the specific conclusion,[2] on the size of the effect measured, and on the statistical confidence in the result.

---

[2]The literature of social sciences contains horror stories of journal editors and others who consider a study worthwhile only if it reaches a statistically significant, positive conclusion; that is, an equally significant rejection of a hypothesis is not considered worthwhile!

Here is the prototype for the analyses treated here: each of the publishable studies consists of repeated measurements of a quantity, say $x$. The number of measurements is in general different in each such study. The null hypothesis is that $x$ is normally distributed, say

$$x \sim \mathcal{N}(\mu, \sigma); \qquad (1)$$

this notation means that the errors in $x$ are normally distributed with mean $\mu$ and variance $\sigma$. The results of the study are reported in terms of a shifted (*i.e.* $\mu$ is subtracted) and renormalized (*i.e.* $\sigma$ is divided out) *standard normal deviate* $Z = \frac{x - \mu}{\sigma}$. The null hypothesis, usually that $\mu$ is zero, one-half, or some other specific value, yields

$$Z \sim \mathcal{N}(0, 1) \qquad (2)$$

This normalization removes the dependence on the number of repeated measurements in the studies. In this step it is of course important that $\sigma$ be well estimated, often a tricky business.

A common procedure for evaluating such studies is to obtain a "$p$-value" from the probability distribution $P(Z)$, and interpret it as providing the statistical significance of the result. The discussion here is confined to this approach because it has been adopted by most researchers in the relevant fields. However, as noted by Sturrock (Sturrock, 1994, Sturrock, 1997) and others (Matthews, 1999, Jefferys, 1990, Jefferys, 1995, Berger and Delampady, 1987, Berger and Sellke, 1987), this procedure may yield incorrect conclusions – usually overestimating the significance of putative anomalous results. The Bayesian methodology is probably the best way to treat publication bias (Givens, Smith, and Tweedie, 1995, Givens, Smith, and Tweedie, 1997, Givens, Smith, and Tweedie, 1997, Biggerstaff, 1995, Biggerstaff, Tweedie, and Mengersen, 1994, Tweedie, Scott, Biggerstaff and Mengersen, 1996).

This section concludes with some historical notes. Of course science has long known the problems of biased samples. The interesting historical note (Petticrew, 1998) nominates an utterance by Diagoras of Melos in 500 BC as the first historical mention of publication bias[3]. See also (Dickersin and Min, 1993) for other early examples of awareness of publication bias.

Publication bias is an important problem in medical studies (*e.g.*, Sacks, H. S., Reitman, D., Chalmers, T. C., & Smith, H., 1983, Persaud, 1996, Allison, Faith, and Gorman, 1996, Laupacis, 1997, Dickersin, 1997, Kleijnen & Knipschild, 1992, Earleywine, 1993, Helfenstein & Steiner, 1994, Faber & Gallœ, 1994), as well as other fields (see Fiske, Rintamäki, & Karvonen, 1998, Bauchau, 1997 for example). The negative conclusions we shall soon reach about the commonly used procedure to deal with this problem yield a discouraging picture of the usefulness of combined analysis in all of these contexts. On the other hand, the application of modern, sophisticated methods such as those listed above (see also Taylor and Tweedie, 1998, Taylor and Tweedie, 1998, Taylor and Tweedie, 1998, Taylor and Tweedie, 1999) is encouraging.

## 3.   The "Fail Safe File Drawer" Calculation

Rosenthal's influential work (Rosenthal, 1979, Rosenthal, 1984, Rosenthal, 1995) is widely used to set limits on the possibility that the file drawer effect is causing a spurious result. One of the clearest descriptions of the overall problem, and certainly the most influential in the social sciences, is (Rosenthal, 1984):

> ... researchers and statisticians have long suspected that the studies
> published in the behavioral sciences are a biased sample of the studies that are actually carried out. ... The extreme view of this problem, the "file drawer problem," is that the journals are filled with the 5% of the studies that show Type I errors, while the file drawers back at the lab are filled with the 95% of the studies that show nonsignificant (*e.g.*, $p > .05$) results.

---

[3] Early Greek sailors who escaped from shipwrecks or were saved from drowning at sea displayed portraits of themselves in a votive temple on the Aegean island of Samothrace, in thanks to Neptune. Answering a claim that these portraits are sure proof that the gods really do intervene in human affairs, Diagoras replied "Yea, but ... where are they painted that are drowned?"

A Type I error is rejection of a true null hypothesis. (Type II is failing to reject a false one.)

This lucid description of the problem is followed by a proposed solution:

In the past, there was very little we could do to assess the net effect of studies tucked away in file drawers that did not make the magic .05 level ... Now, however, although no definitive solution to the problem is available, we can establish reasonable boundaries on the problem and estimate the degree of damage to any research conclusions that could be done by the file drawer problem. The fundamental idea in coping with the file drawer problem is simply to calculate the number of studies **averaging null results** that must be in the file drawers before the overall probability of a Type I error can be just brought to any desired level of significance, say $p = .05$. This number of filed studies, or the tolerance for future null results, is then evaluated for whether such a tolerance level is small enough to threaten the overall conclusion drawn by the reviewer. If the overall level of significance of the research review will be brought down to the level of *just significant* by the addition of just a few more **null results**, the finding is not resistant to the file drawer threat.

The *italic* emphasis is original; I have indicated what I believe is the fundamental flaw in reasoning with **boldface**.

By its very definition, **the file drawer is a biased sample**. In the nominal example given, it is the 95% of the studies that have 5% or greater chance of being statistical fluctuations. The mean $Z$-value in this subsample is not zero, but instead

$$\bar{Z}_{filed} = -\frac{1}{\sqrt{(2\pi)}} \frac{e^{-\frac{1}{2}Z_0^2}}{.95} = -0.1085, \tag{3}$$

where $Z_0$ is the value corresponding to the adopted signficiance threshold. As we will see below, Rosenthal's analysis explicitly assumes that $\bar{Z}_{filed} = 0$ for the file drawer sample. Because this assumption contradicts the essence of the file drawer effect, the quantitative results are incorrect.

We now recapitulate the analysis given in (Rosenthal, 1984), using slightly different notation. For convenience and consistency with the literature, we refer to this as the fail safe file drawer, or **FSFD**, analysis. The basic context is a specific collection of published studies having a combined $Z$ that is deemed to be significant – that is, the probability that the $Z$ value is due to a statistical fluctuation is below some threshold, say 0.05. The question Rosenthal seeks to answer is, How many members of a hypothetical set of unpublished studies have to be added to this collection in order to bring the mean $Z$ down to a level considered insignificant. As argued elsewhere, this does not mirror publication bias, but this is the problem addressed.

Let $N_{pub}$ be the number studies combined in the analysis of the published literature (Rosenthal's $K$), and $N_{filed}$ be the number of studies that are unpublished, for whatever reason (Rosenthal's $X$). Then $N = N_{pub} + N_{filed}$ is the total number of studies carried out.

The basic relation [Equation (5.16) of (Rosenthal, 1984)] is,

$$1.645 = \frac{N_{pub}\bar{Z}_{pub}}{\sqrt{N}} \tag{4}$$

This is presumably derived from

$$1.645 = \frac{N_{pub}\bar{Z}_{pub} + N_{filed}\bar{Z}_{filed}}{\sqrt{N}} \tag{5}$$

*i.e.* an application of his *method of adding weighted Z's* as in Equation [5.5] of (Rosenthal, 1984), by setting the standard normal deviate of the file drawer, $\bar{Z}_{filed} = 0$. This is incorrect for a biased file drawer. Equation (4) can be rearranged to give

$$\frac{N}{N_{pub}} = \frac{N_{pub}\bar{Z}_{pub}^2}{2.706} \tag{6}$$

which is the equation used widely in the literature, and throughout this paper, to compute **FSFD** estimates.

What fundamentally went wrong in this analysis and why has it survived uncriticized for so long? First, it is simply the case that the notion of *null, or insignificant results* is easily confused with $\bar{Z}_{filed} = 0$. While the latter implies the former, the former (which is true, in a sense, for the file drawer) of course, does not imply the latter.

Second, the logic behind the **FSFD** is quite seductive. Pouring a flood of insignificant studies – with normally distributed $Z$'s – into the published sample until the putative effect submerges into insignificance is a neat idea. What does it mean though? It is indeed a possible measure of the statistical fragility of the result obtained from the published sample. On the other hand, there are much better and more direct ways of assessing statistical significance, so that I do not believe that **FSFD** should be used even in this fashion.

Third, I believe some workers have mechanically calculated **FSFD** results, found that it justified their analysis or otherwise confirmed their beliefs, and were therefore not inclined to look for errors or examine the method critically. A simple thought experiment makes the threat of a publication bias with $N_{filed}$ on the same order as $N_{pub}$ clear: Construct a putative file drawer sample by multiplying all published $Z$ values by $-1$; then the total sample then has $\bar{Z}$ exactly zero, no matter what.

The two-sided case is often raised in this context. That is to say, publication bias might mean that studies with either $Z >> 0$ or $Z << 0$ are published in preference to those with small $| Z |$. This situation could be discussed, but it is a red herring in the current context (see *e.g.* Iyengar & Greenhouse, 1988). Further, none of the empirical distributions I have seen published show any hint of the bimodality that would result from this effect, including those in (Radin, 1997).

In addition, I was first puzzled by statements such as that **FSFD** analysis assesses "tolerance for future null results." The file drawer effect is something that has already happened by the time one is preparing a combined analysis. I concluded that such expressions are a kind of metaphor for what would happen if the studies in the file drawers were suddenly made available for analysis. But even if this imagined event were to happen, one would still be left with explaining how a biased sample was culled from the total sample. It seems to me that whether the result of opening this Pandora's file drawer is to dilute the putative effect into insignificance, or not, is of no explanatory value in this context – even if the calculation were correct.

In any case, the bottom line is that the widespread use of the **FSFD** to conclude that various statistical analyses are robust against the file drawer effect is wrong, because the underlying calculation is based on an inapplicable premise. Other critical comments have been published, including (Berlin, 1998, Sharpe, D., 1997)). Most important is the paper (Iyengar & Greenhouse, 1988), which makes the same point, but further provides an analysis that explicitly accounts for the bias (for the case of the truncated selection functions discussed in the next section). These authors note that their formulation "always yields a smaller estimate of the fail-safe sample size than does" Rosenthal's.

## 4.   **Statistical Models**

This section presents a specific model for combined analysis and the file drawer effect operating on it. Figure 1 shows the distribution of $N = 1000$ samples from a

Fig. 1.— Histogram corresponding to the null hypothesis: normal distribution of $Z$ values from 1000 independent studies. Those 5% with $Z \geq 1.645$ are published (open bars), the remainder "tucked away in file drawers" – *i.e.*, unpublished (solid bars). Empirical and exact values of $Z$ [*cf.* Equation (14)] are indicated for the filed and published sets.

normal distribution with zero mean and unit variance, namely

$$G(Z) = \frac{N}{\sqrt{2\pi}} e^{-\frac{1}{2}Z^2} \qquad (7)$$

Publication bias means that the probability of completion of the entire process detailed in Section 2 is a function of the study's reported $Z$. Note that we are not assuming this is the only thing that it depends on. We use the notation

$$S(Z) = \text{publication probability} \tag{8}$$

for the selection function, where $0 \leq S(Z) \leq 1$. Note $S(Z)$ is **not** a probability distribution over $Z$, and for example its $Z$-integral is not constrained to be 1.

This model captures what almost all authors describe as publication bias (see *e.g.* Hedges, 1992, Iyengar & Greenhouse, 1988). In view of the complexity of the full publication process (see Section 2), it is unlikely that its psychological and sociological factors can be understood and accurately modeled. I therefore regard the function $S$ as unknowable. The approach taken here is to study the generic behavior of plausible quantitative models, with no claim to having accurate or detailed representations of actual selection functions.

Taken with the distribution $G(Z)$ in Equation (7), an assumed $S(Z)$ immediately yields three useful quantities: the number of studies published

$$N_{pub} = \int_{-\infty}^{\infty} G(Z)S(Z)dZ \ , \tag{9}$$

the number consigned to the file drawer

$$N_{filed} = N - N_{pub} \ , \tag{10}$$

and the expected value of the standard normal deviate $Z$, averaged over the published values

$$\bar{Z}_{pub} = \frac{\int_{-\infty}^{\infty} ZG(Z)S(Z)dZ}{\int_{-\infty}^{\infty} G(Z)S(Z)dZ} = \frac{\int_{-\infty}^{\infty} ZG(Z)S(Z)dZ}{N_{pub}} \tag{11}$$

The denominator in the last equation takes into account the reduced size of the collection of published studies ($N_{pub}$) relative to the set of all studies performed ($N$). Absent any publication bias [*i.e.* $S(Z) \equiv 1$] $\bar{Z}_{pub} = 0$ from the symmetry of the Gaussian, as expected.

### 4.1.   Cut Off Selection Functions

Consider first the following choice for the selection function:

$$S(Z) = \left\{ \begin{array}{ll} 0 & \text{for } Z < Z_0 \\ 1 & \text{for } Z \geq Z_0 \end{array} \right. , \tag{12}$$

where in principle $Z_0$ can have any value, but small positive numbers are of most practical interest. That is to say, studies which find an effect at or above the significance level corresponding to $Z_0$ are always published, while those that don't never are.

Putting this into the above general expressions Equations (9) and (11) gives

$$N_{pub} = \frac{N}{\sqrt{2\pi}} \int_{Z_0}^{\infty} e^{-\frac{1}{2}Z^2} dZ \; = \frac{N}{2} \, \text{erfc}(\frac{Z_0}{\sqrt{2}}), \tag{13}$$

where **erfc** is the *complementary error function*. The mean $Z$ for these published studies is

$$\bar{Z}_{pub} = \frac{N}{N_{pub}} \frac{1}{\sqrt{2\pi}} \int_{Z_0}^{\infty} Z e^{-\frac{1}{2}Z^2} dZ = \sqrt{(\frac{2}{\pi})} \frac{e^{-\frac{1}{2}Z_0^2}}{\text{erfc}(\frac{Z_0}{\sqrt{2}})} \; . \tag{14}$$

Consider the special value $Z_0 = 1.645$, corresponding to the classical 95% confidence level and obtained by solving for $Z$ the equation

$$p = \frac{1}{2}\text{erfc}(\frac{Z}{\sqrt{2}}) \tag{15}$$

(with $p = .05$). This case is frequently taken as a prototype for the file drawer (Rosenthal, 1979, Iyengar & Greenhouse, 1988). Equation (13) gives $N_{pub} = 0.05N$; that is (by design) 5% of the studies are published and 95% are not. Equation (14) gives $\bar{Z}_{pub} = 2.0622$.

Let the total number of studies, be $N = 1000$: *i.e.*, one thousand studies, of which 50 are published and 950 are not. We have chosen this large value for $N$, here and in Figure 1, for illustration, not realism. For the 50 published studies, the combined $Z$, namely $\sqrt{50}\bar{Z}_{pub}$, in Equation(15) gives an infinitesimal $p$-value, $\approx 10^{-48}$ – highly supportive of rejecting the null hypothesis. The **FSFD** estimate of the ratio of filed to published experiments [see Equation(6)] is about 78 for this case, an overestimate, by a factor of around 4, of the true value of 19. The formula of Iyengar and Greenhouse discussed above gives 11.4 for the same ratio, an underestimate by a factor of about 1.7.

Fig. 2.— Plot of the fraction of the total number of studies that are published (thick solid line) and filed (thick dashed line) in the model given by Equation (12). The **FSFD** predictions for the filed fraction are shown as a series of thin dashed lines, labled by the value of $N$, the total number of studies carried out.

Finally, Figure 2 shows the behavior of the filed and published fractions, as a function of $Z_0$, including the fraction of the studies in the file drawer predicted by **FSFD** analysis, given by Equation (6) above. Two things are evident from this comparison:

- The **FSFD** prediction is a strong function of $N$, whereas the true filed fraction is independent of $N$.

- the **FSFD** values are quite far from the truth, except accidentally at a few special values of $N$ and $Z_0$.

## 4.2.  Step Selection Functions

Since it is zero over a significant range, the selection function considered in the previous section may be too extreme. The following generalization allows any value between zero and one for small $Z$:

$$S(Z) = \begin{cases} S_0 & \text{for } Z < Z_0 \\ 1 & \text{for } Z \geq Z_0 \end{cases} \quad (0 \leq S_0 \leq 1) \tag{16}$$

Much as before, direct integration yields

$$N_{pub} = N[S_0 + (1 - S_0)\frac{1}{2} \text{ erfc}(\frac{Z_0}{\sqrt{2}})], \tag{17}$$

and

$$\bar{Z}_{pub} = (1 - S_0)\sqrt{(\frac{2}{\pi})}\frac{e^{-\frac{1}{2}Z_0^2}}{\text{erfc}(\frac{Z_0}{\sqrt{2}})} \tag{18}$$

Equation (17) directly determines the ratio of the number of filed studies to the number of those published, under the given selection function. The value of $\bar{Z}_{pub}$ is the quantity that would be used to (incorrectly) reject the hypothesis "No effect is present; the sample is drawn from a normal distribution."

Figure 3 shows this function; The basic result is that almost all of the relevant part of the $S_0 - Z_0$ plane corresponds to a rather small number of unpublished studies. In particular, $R \gg 1$ in only a small region, namely where simultaneously $S_0 \sim 0$ and $Z_0 >> 0$. The behavior of $\bar{Z}_{pub}$ is quite simple: roughly speaking

$$\bar{Z}_{pub} \approx (1 - S_0)g(Z_0) \tag{19}$$

where $g(Z_0)$ is a function on the order of unity or larger for all $Z_0 > 0$. Hence the bias brought about by even a very small file drawer is large unless $S_0 \approx 1$ (to be expected, because for such values of $S_0$ the selection function is almost flat).

Fig. 3.— This figure shows the dependence of $R \equiv N_{filed}/N_{pub}$, on the two parameters of the step selection function. Contours of $log_{10}R$ are indicated. $R$ is large only for a tiny region region at the bottom right of this diagram.

### 4.3.   Smooth Selection Functions

It might be objected that the selection functions considered here are unreal in that they have a discrete step, at $Z_0$. What matters here however, are integrals over $Z$, which do not have any pathological behavior in the presence of such steps. Nevertheless I experimented with some smooth selection functions and found results that are completely consistent with the conclusions reached in the previous sections for step-function choices.

## 5.  Conclusions

Based on the models considered here, we conclude that:

- apparently significant, but actually spurious, results can arise from publication bias, with only a modest number of unpublished studies

- the widely used Fail Safe File Drawer (**FSFD**) analysis is irrelevant, because it treats the inherently biased file drawer as unbiased, and gives grossly wrong estimates of the size of the file drawer

- statistical combination of studies from the literature can be trusted to be unbiased only if there is reason to believe that there are essentially no unpublished studies (almost never the case!)

It is hoped that these results will discourage combined ("meta") analyses based on selection from published literature, but encourage methodology to control publication bias, such as the establishment of registries (to try to render the probability of publication unity once a study is proposed and accepted in the registry).

The best prospects for establishing conditions under which combined analysis might be reasonable even in the face of possible publication bias, seem to lie in a fully Bayesian treatment of this problem (Sturrock, 1994, Sturrock, 1997, Givens, Smith, and Tweedie, 1997). It is possible that the approach discussed in (Radin & Nelson, 1989) can lead to improved treatment of publication bias, but one must be cautious when experimenting with *ad hoc* distributions and be aware that small errors in fitting the tail of a distribution can be multiplied by extrapolation to the body of the distribution.

Finally, I agree with the Referee, Prof. M. J. Bayarri, who wants to go even further, stating: " ... the Publication Bias effect should be taken into account even when making conclusions based on a *single* published experiment." Certainly the bias mechanism described in Section 2 can determine which single paper on a given topic gets published. To say the same thing, $N_{pub} = 1$ is a perfectly legitimate case in my analysis. Indeed, one can think of several psychological/sociological factors that could militate against publication of a second paper on a given experiment.

I am especially grateful to Peter Sturrock for guidance and comments in the course of this work. This work was presented at the Peter A. Sturrock Symposium, held on March 20, 1999, at Stanford University. I thank Kevin Zahnle, Aaron Barnes, Ingram Olkin, Ed May, and Bill Jefferys for valuable suggestions. I am grateful to the Referee for making helpful comments and pointing out several additional references. I thank Jordan Gruber for calling my attention to the book *The Conscious Universe* (Radin, 1997), and its author, Dean Radin, for helpful discussions. None of these statements are meant to imply that any of the persons mentioned agrees with the conclusions expressed here.

# REFERENCES

Allison, D. B., Faith, M. S., and Gorman, B.S. (1996). Publication bias in obesity treatment trials? *International Journal of Obesity*, 20, 10, 931-937.

> Discussion of publication bias in obesity treatment studies, mentioning the **FSFD** method.

Bauchau, V. (1997). Is there a "file drawer problem" in biological research. *OIKOS*, 19, 2, 407-409.

> Concludes that the extent of the file drawer problem in biology is an open question.

Bayarri, M. J. and Berger, J. O. (1998). *The Annals of Statistics*, 26, 645-659.

> Nonparametic Bayesian analysis.

Bayarri, M. J. and DeGroot, M. H. (1987). Bayesian analysis of selection models, *The Statistician*, 36, 137-146.

Bayarri, M. J. and DeGroot, M. H. (1990). The Analysis of Published Significant Results, in Rassegna di Metodi Statistici ed Applicazioni.

> This and the previous paper are excellent discussions of the overall problem and the Bayesian approach to it; both include many other references.

Berger, J. O., and Delampady, M. (1987), Testing Precise Hypotheses. *Statistical Science*, 2, 317-352 (with discussion).

> This and the following paper are recommended by the Referee as key criticisms of $p$-values as adequate measures of evidence against the null hypothesis; he also points out the existence of several relevant discussion papers by Berger, Bayarri, and others on the WWW site of the Duke University Institute of Statistics and Decision Studies, at `http://www.stat.duke.edu/`.

Berger, J. O., and Sellke, T. (1987), Testing a point null hypothesis: the Irreconciliability of $p$-values and evidence. *J. American Statistical Association*, 82, 112-122.

Berlin, J. A. (1998). Publication Bias in Clinical Research: Outcome of Projects Submitted to Ethics Committees. *Clinical Trials in Infant Nutrition*, Eds. Perman, J. A. and Rey, J., Nestlé Nutrition Workshop Series, Vol. 40, Nestec Ltd., Vevet-Lippincott-Raven Publishers, Philadelphia.

> Interesting overview of the subject of publication bias. "The existence of publication bias is clearly demonstrated by the studies described." (That is, various health and nutrition studies.) A clear restatement of the **FSFD** analysis is followed by the statement that "this approach is of limited utility for two reasons: first, because it uses only $Z$ statistics and ignores quantitative estimates of effects (for example odds ratios); and second, because the assumption that all the unpublished studies have a $Z$ statistic of exactly zero is unrealistic." The paper is followed by an interesting, free-wheeling discussion that touches on publication in obscure journals (something that could be added to the discussion in Section 2), something called *reference bias*, electronic journals, and registries.

Biggerstaff, B. J. (1995). Random Effects Methods in Meta-Analysis with Application in Epidemiology. Colorado State University Ph. D. thesis.

Comprehensive discussion of a number of issues in combined studies, including random effects and Bayesian methods, with application to environmental tobacco smoke and lung cancer studies.

Biggerstaff, B. J., Tweedie, R. L., and Mengersen, K. L. (1994). Passive Smoking in the Workplace: Classical and Bayesian Meta-analyses. *Int. Arch. Occupational and Environmental Health*, to appear.

Discussion and comparison of classical and Bayesian approaches to combined analysis.

Buckheit, J., and Donoho, D. (1995). WaveLab and Reproducible Research. in *Wavelets and Statistics*, Antoniadis and Oppenheim eds., Springer-Verlag, Lecture Notes in Statistics No. 103.

Csada, R. D., James, P. C., & Espie, R. H. M. (1996). The "file drawer problem" of non-significant results: does it apply to biological research? *OIKOS*, 76, 3, 591-593.

An empirical study that quantitatively supports the suspicion that publication bias is important in biology.

Claerbout, J., Schwab, and Karrenbach, M. (1999). various documents at `http://sepwww.stanford.edu/research/redoc/`.

Dickersin, K. (1997). How Important Is Publication Bias? A Synthesis of Available Data. *AIDS Education and Prevention*, 9A, 15-21.

Empirical study that reaches the conclusion that "publication is dramatically influenced by the direction and strength of research findings." Found that editors were less enthusiastic about including unpublished studies in combined analysis.

Dickersin, K., and Min, Y. (1993). Publication bias: the problem that won't go away. *Ann. NY Acad. Sci.*, Vol. 703, 135-146.

Earleywine, M. (1993). The File Drawer Problem in the Meta-Analysis of Subjective Responses to Alcohol. *Am. J. Psychiatry*, 150, 9, 1435-1436.

Applies the **FSFD** calculation to a combined analysis of 10 published studies, by V. Pollock, of subjective responses to alcohol and risk for alcoholism. The resulting **FSFD** $N$ is 26. In a response, Pollock agrees with the (dubious) conclusion that this means the analysis can be considered resistant to the file drawer problem.

Faber, J. & Gallœ, A. M. (1994). Changes in bone mass during prolonged subclinical hyperthyroidism due to L-thyroxine treatment: a meta-analysis. *European Journal of Endocrinology*, 130, 350-356.

Analysis combining 13 studies of a thyroid treatment; the **FSFD** calculation yielded 18 as the "fail-safe $N$," which seems to be taken as evidence of the resistance of the results of the analysis to publication bias.

Fiske, P., Rintamäki, P. T., & Karvonen, E. (1998). Mating success in lekking males: a meta-analysis. *Behavioral Ecology*, 9, 4, 328-338.

Combined analysis of zoological data. "When we used this [the **FSFD**] method on our sigficant effects, the number of 'hidden studies' needed to change our results ranged from 15 ... to 3183 ..., showing that most of our results are quite robust." Robust against including a set of unbiased unpublished results, but not against the relevant biased set.

Givens, G. H., Smith, D. D., and Tweedie, R. L. (1995). Estimating and Adjusting for Publication Bias Using Data Augmentation in Bayesian Meta-Analysis, Technical Report 95/31, Department of Statistics, Colorado State University.

A Bayesian analysis of publication bias, using the *data augmentation principle*, with simulations and application to data from 35 studies of the relation between lung cancer and spousal exposure to environmental tobacco smoke.

Givens, G. H., Smith, D. D., and Tweedie, R. L. (1997). Publication Bias in Meta-Analysis: A Bayesian Data-Augmentation Approach to Account for Issues Exemplified in the Passive Smoking Debate. *Statistical Science*, Vol. **12**, pp. 221-250.

A Bayesian analysis of publication bias, based on choosing a model for the selection function and marginalizing over the parameters of the model.

LaFleur, B., Taylor, S., Smith, D. D., Tweedie. R. L. (1997). Bayesian Assessment of Publication Bias in Meta-Analysis of Cervical Cancer and Oral Contraceptives, University of Colorado reprint?

Application of a Bayesian method for assessing publication bias to studies of the possible effects of oral contraceptive use on the incidence of cervical cancer. They conclude that publication bias, probably caused by the explicit disregard of 'low quality' studies, yielded a spurious statistical connection between oral contraceptive use and the incidence of cervical cancer in a previous study.

Hedges, L. V. and Olkin, I (1985). *Statistical Methods for Meta-Analysis*, Academic Press.

A classic, with an extensive discussion (Chapter 14) of the effect of publication bias for one or more experiments.

Hedges, L. V. (1992). Modeling Publication Selection Effects in Meta-Analysis. *Statistical Science*, 7, 2, 246-255.

Interesting overview, with comments on the nature of the publishing process and results that are "not statistically significant." Introduces a quantitative model for publication bias that is much like that given here, and offers a quantitative test for the presence of publication bias.

Helfenstein, U., Steiner, M. (1994). Fluoride varnishes (Duraphat): A meta-analysis. *Community Dentistry and Oral Epidemiology*, 22, 1-5.

The authors apply the **FSFD** method to a combined analysis of studies designed to detect the cavity preventive effect of a fluoride varnish called Duraphat. Application of the **FSFD** method yields the conclusion (unwarranted, based on my conclusions) that "It is very unlikely that underreporting of non-significant results could reverse the conclusion into an overall null-result."

Iyengar, S., and Greenhouse, J.B. (1988). Selection Models and the File-Drawer Problem, *Statistical Science*, **3**, 109-135.

An excellent overview and nearly definitive study using much the same selection function approach as used here, reaching more or less the same critical conclusions. In addition, these author's Equation (4) offers a presumably improved basis for the **FSFD** estimate. The paper is followed by extensive comments by Larry Hedges, by Robert Rosenthal and Donald Rubin, by Nan Laird, G. Patil and C. Taillie, by M. Bayarri, by C. Radhakrishna Rao, and by William DuMouchel – all followed by a rejoinder by Igengar and Greenhouse. Much of this discussion seems to evade the simple issue raised here.

Jefferys, W. H. (1990). Bayesian Analysis of Random Event Generator Data. *Journal of Scientific Exploration*. 4, 153-169.

This excellent paper criticizes the classical use and interpretation of "p-values" in the analysis of data from psychic random number generator experiments, and argues for a Bayesian alternative.

Jefferys, W. H. (1995). *Journal of Scientific Exploration*. Letters to the Editor, 9, 121-122 and 595-597.

> Incisive comments on the meaning of "p-values".

Kleijnen J., & Knipschild, P. (1992). Review Articles and Publication Bias. *Arzneim.-Forsch./Drug Res.*, 42, 5, 587-591.

> Overview of publication bias in drug research.

Laupacis, A. (1997). Methodological Studies of Systematic Reviews: Is There Publication Bias? *Arch. INtern. Med.*. 157, 357-358.

> Brief suggestion that publication bias should be considered in medical research.

Matthews, Robert A. J. (1999). Significance Levels for the Assessment of Anomalous Phenomena. *Journal of Scientific Exploration*, Vol. **13**, pp. 1-7.

Persaud, Rajendra (1996). Studies of Ventricular Enlargement. *Arch. Gen. Psychiatry*, Vol. **53**, December, 1996, p. 1165.

> Criticism of a combined study of ventricular enlargement in patients with mood disorders, for misinterpretation of the results of a **FSFD** calculation. In the case at issue, a "fail-safe N" of 10 results for an analysis with 11 studies. The closing statement ("A few studies that are highly significant, even when their combined P value is significant, may well be misleading because only a few unpublished, or yet to be published, studies could change the combined significant result to a nonsignificant result.") seems to capture what people must have in mind for the significance of the **FSFD** computation.

Petticrew, M. (1998). Diagoras of Melos (500 BC): an early analyst of publication bias. *The Lancet*, 352, 1558.

Radin, D. I. (1997). *The Conscious Universe: The Scientific Truth of Psychic Phenomena*, Harper Edge.

> A number of combined studies are displayed and offered as evidence for the reality of psychic phenomena. The possibility that these results are spurious and due to publication bias is considered, and then rejected because the FSFD calculation yields huges values for the putative file drawer. It is my opinion that, for the reasons described here, these values are meaningless and that publication bias may well be responsible for the positive results derived from combined studies.

Radin, D. I. and Nelson, R. D. (1997). Evidence for Consciousness-Related Anomalies in Random Physical Systems. *Foundations of Physics*, 19, 12, 1499-1514.

Combined study of 597 experimental random number generator studies. An innovative method of dealing with the file drawer effect is used, based on fitting the high-end tail of the $Z$ distribution. I question the use of an exponential model for this tail, justified by the comment that it is used "to simulate the effect of skew or kurtosis in producing the disproportionately long positive tail."

Riniolo, T. C. (1997). Publication Bias: A Computer-Assisted Demonstration of Excluding Nonsignificant Results From Research Interpretation. *Teaching of Psychology*, 24, 4, 279-282.

Describes an educational software system that allows students to perform statistical experiments to study the effects of publication bias.

Rosenthal, R. (1978). Combining Results of Independent Studies. *Psychological Bulletin*, 85, 1, 185-193.

Comprehensive and influential (158 references) analysis of the statistics of combined studies.

Rosenthal, R. (1979). The 'File Drawer Problem' and Tolerance for Null Results. *Psychological Bulletin*, Vol. **86**, No. 3, pp. 638-641.

This is the standard reference, cited in almost all applied research in which the file drawer effect is at issue.

Rosenthal, R. (1984). *Meta-Analytic Procedures for Social Research* Applied Social Research Methods Series, Volume 6, Sage Publications: Newbury Park.

The treatment of the file-drawer effect, in Chapter 5, Section II.B, is essentially identical to that in (Rosenthal, 1979).

Rosenthal, R. (1990). Replication in Behavioral Research. in *Handbook of replication research in the behavioral and social sciences*, special issue of *Journal of Social Behavior and Personality* 5, 4, 1-30.

Overview of the issue of replication, including a brief discussion of the **FSFD** approach.

Rosenthal, R. (1995). Writing Meta-Analytic Reviews. *Psychological Bulletin*, Vol. **118**, No. 2, pp. 183-192.

A broad overview of combined studies.

Rosenthal, R. (1991). Meta-Analysis: A Review. *Psychosomatic Medicine*, Vol. **53**, pp. 247-271.

Comprehensive overview of various procedural and statistical issues in analysis of combined data. There is a section on the File Drawer Problem (p. 260), more or less repeating the analysis in (Rosenthal, 1979), with an illustrative example.

Sacks, H. S., Reitman, D., Chalmers, T. C., & Smith, H., Jr. (1983). The Effect of Unpublished Studies (The File Drawer Problem) on Decisions about Therapeutic Efficacy. *Clinical Research*, 31, 2, 236.

This abstract applies the FSFD calculation to a number of medical studies, dealing with anticoagulants for acute myocardial infarction, and other treatments. The computed file drawer sizes are not large in most cases, but the concluding statement "This is a useful measure of the strength of the published evidence on any question," is wrong, for reasons given in this paper.

Sharpe, D. (1997). Of Apples and Oranges, File Drawers and Garbage: Why Validity Issues in Meta-Analysis Will Not Go Away, *Clinical Psychology Review*, 17, 8, 881-901.

This is an extensive review of a number of problems in combined analysis, and includes an extensive critical discussion of the **FSFD** approach, commenting that the rule behind it is arbitrary – but falling short of saying that the method is wrong.

Silvertown, J. & McConway, K. J. (1997). Does "publication bias" lead to biased science? *OIKOS*, 79, 1, 167-168.

Short commentary on publication bias in biology. The authors mention several techniques for studying the presence and scope of publication bias, and generally downplay the seriousness of the problem

Smith, D. D., Givens, G. H., and Tweedie, R. L. (1999) Adjustment for Publication and Quality Bias in Bayesian Meta-Analysis, Colorado Staue University reprint?

Development of a data augmentation technique to assess potential publication bias.

Stanton, M. D. & Shadish, W. R. (1997) Outcome, Attrition, and Family-Couples Treatment for Drug Abuse: A Meta-Analysis and Review of the Controlled, Comparative Studies. *Psychological Bulletin*, 122. 2, 170-191

Combined analysis of a large number of psychological studies; mentions the **FSFD** problem but does not appear to do anything about it.

Sturrock, P. A. (1994). Applied Scientific Inference. *Journal of Scientific Exploration*, Vol. **8**, No. 4, pp. 491-508.

A clear and incisive discussion of a Bayesian formalism for combining judgements in the evaluation of scientific hypotheses from empirical data, such as in investigations of anomalous phenomena.

Sturrock, P. A. (1997). A Bayesian Maximum-Entropy Approach to Hypothesis Testing, for Application to RNG and Similar Experiments. *Journal of Scientific Exploration*, Vol. **11**, No. 2, pp. x-xx.

> Cogent critique of the standard "*p*-value" test in applied scientific inference. Shows that the Bayesian formalism, based on a clear statement and enumeration of the relevant hypotheses, is superior to and different from "*p*-value" analysis.

Taylor, S., and Tweedie, R. (1998). A non-parametric 'trim and fill' method of assessing publication bias in meta-analysis, U. Colorado preprint?

Taylor, S., and Tweedie, R. (1998). Trim and fill: A simple method of assessing publication bias in meta-analysis, U. Colorado preprint?

Taylor, S., and Tweedie, R. (1998). Trim and fill: A simple Funnel Plot Based Method of Testing and Adjusting for Publication Bias in Meta-Analysis, U. Colorado preprint.

> Discussion and application of the trim and fill method to a number of combined studies.

Taylor, S., and Tweedie, R. (1999). Practical Estimates of the Effect of Publication Bias in Meta-Analysis, U. Colorado preprint.

Tweedie, R. L., Scott, D. J., Biggerstaff, B.J., and Mengersen, K.L. (1996). "Bayesian meta-analysis, with application to studies of ETS and lung cancer. *Lung Cancer*, **14**, S171-194, Suppl. 1.

> A comprehensive analysis of publication bias, comparing a Markov chain Monte Carlo technique for implementing Bayesian hieratchical models with *random effects* models and other classical methods.

---